\documentclass{article}
\usepackage{graphicx} 
\usepackage{todonotes}
\usepackage{listings}
\usepackage{authblk}
\usepackage{amsmath}
\usepackage{hyperref}

\usepackage{color}

\definecolor{dkgreen}{rgb}{0,0.6,0}
\definecolor{gray}{rgb}{0.5,0.5,0.5}
\definecolor{mauve}{rgb}{0.58,0,0.82}

\title{Enabling the Verification and Formalization of Hybrid Quantum-Classical Computing with OpenQASM 3.0 compatible QASM-TS 2.0}

\author[1]{Sean Kim\thanks{skim658@gwu.edu}}
\author[2]{Marcus Edwards\thanks{msedward@student.ubc.ca}}
\affil[1]{The Department of Biochemistry \& Molecular Medicine,
Washington DC 20037, United States of America}
\affil[2]{Electrical and Computer Engineering,
University of British Columbia, Vancouver, BC, CA}

\date{September 2025}

\begin{document}

\maketitle

\section{Summary}

The unique features of the hybrid quantum-classical computing model implied by the specification of OpenQASM 3.0 motivate new approaches to quantum program verification. We implement and thoroughly test a QASM 3.0 parser in TypeScript to enable implementations of verification and validation software, compilers, and more. We aim to help the community to formalize the logic of hybrid quantum-classical computing by providing tools that may help with such efforts.

\subsection{Top Level Abstractions}

The parser implements recursive descent parsing with support for expression precedence and type checking. At a high level, the parsing can be split into three logical sections: expression parsing, quantum-specific parsing, and classical parsing.

Expression parsing forms the foundation for both quantum and classical parsing by breaking down expressions into their constituent parts and rebuilding them according to the mathematical and operator syntax defined in the OpenQASM 3.0 official grammar \cite{noauthor_qasm_grammar_nodate}. Starting with basic elements like numbers, variables, and operators, the parser constructs an abstract syntax tree that reflects proper operator precedence and nesting. This process handles not only mathematical operations but also array accesses, function calls, and parameter lists, creating a structured representation that maintains the logical relationships between all parts of the expression.

Quantum operation parsing manages the core quantum computing elements of QASM, ensuring that quantum operations are syntactically correct. The parser maintains strict tracking of quantum resources through its gates, standardGates, and customGates sets, validating that each quantum operation references only properly defined gates and qubits. It processes quantum register declarations, custom gate definitions, gate applications (including gate modifiers), measurement operations, and timing-critical operations like barriers and delays. Each quantum operation is validated in its context.

The classical parsing processes classical variable declarations with strict type checking, function definitions with parameter and return type validation, control flow structures like conditional statements and loops, and array operations. The parser maintains separate tracking for classical and quantum resources while ensuring they can interact in well-defined ways. This enables QASM programs to express complex quantum algorithms that require classical processing while maintaining type safety and operational validity.

Our parser produces a strongly-typed AST that captures the full structure of QASM programs and is designed to enable subsequent semantic analysis.

\section{Statement of Need}

The OpenQASM 3.0 type system supports classical and quantum types as well as functions which can be used to specify hybrid quantum-classical programs.

There is need for formal analysis and verification of hybrid quantum classical programs and we argue that mathematical frameworks and software frameworks are needed to address this gap.

Hand-in-hand with formalization efforts are the development of community standards for quantum programming. The standardization of quantum computer programming is ongoing \cite{di_matteo_abstraction_2024, Cross_2022} and relies significantly on open source software and frameworks including some released only last year \cite{seidel2024qrispframeworkcompilablehighlevel, qdmi}. Some community standards such as the 2022 Open Quantum Assembly (OpenQASM) 3.0 specification boast typing as well as interoperability and portability between quantum systems of different types. However, others such as the QUASAR instruction set architecture assume some backend details such as a classical co-processor to complement the Quantum Processing Unit (QPU) \cite{shammah_open_2024}. To what extent the classical surrounds of a quantum processing unit should be assumed or specified and at what part of the stack is an open question. This "piping" can leverage many classical programming paradigms including web technology. What we refer to here is distinct from cloud quantum computing which simply offers a web-accessible front-end to users of quantum computers. Instead, we are interested in parts of the programming model itself, such as pieces of the compile toolchain (compilers, transpilers, assemblers, noise profilers, schedulers) that are implemented using web technology. An example of this is Quantinuum's QEC decoder toolkit, which uses a WebAssembly (WASM) virtual machine (WAVM) as a real-time classical compute environment for QEC decoding \cite{noauthor_qec_nodate}. Other examples include our ports of Quantum Assembly (QASM), Quantum Macro Assembler (QMASM), and Blackbrid to TypeScript \cite{edwards_three_2023}.

An important part of standardization is verification. Our typed OpenQASM 3.0 parser implements a system that infers types from QASM syntax. This opens the door to the type-based formal verification of QASM code. A body of work exists regarding the verification of quantum software, and it is summarized in \cite{exman_verification_2024}.

The primary future direction that we see is the development of verification tools such as static analysis tools based on QASM-TS in the vein of QChecker \cite{zhao_qchecker_2023}. This could be complemented by a formal type theory of OpenQASM 3.0.

Virtually every quantum computing company has provided access through a hybrid cloud. This demands that parts of the stack be implemented in web technology, and we argue that it is optimal in a sense to use technology that is designed for this environment when we find ourselves working in a hybrid quantum / classical cloud. We suggest that a closer marriage of open source efforts to the inherently web-based stack supporting existing quantum computing offerings is desirable.

We note that Osaka University's open-source quantum computer operating system project "Oqtopus" already depends on and makes use of Qasm-ts \cite{osaka_2025} and thank the Oqtopus team for their interest in our work.

\section{Outcomes}

Our comparative analysis focused on two promiment OpenQASM 3.0 parsers: Qiskit's Python ANLTR-based reference implementation and Qiskit's experimental Rust parser.

\subsection{Performance Benchmarking}

\noindent \begin{tabular}{ |p{3cm}|p{3cm}|p{3cm}|p{3cm}|  }
 \hline
 \multicolumn{4}{|c|}{Benchmark Results} \\
 \hline
 Result& ANTLR Parser&Rust Parser&Qasm-ts\\
 \hline
 Success Rate&100\% (11/11 files)&18.2\% (2/11 files)&100\% (11/11 files)\\
 Average Time&8.53 ms&0.59 ms&0.90 ms\\
 Min Time&1.93 ms&0.55 ms&0.36 ms\\
 Max Time&30.54 ms&0.62 ms&3.68 ms\\
 \hline
\end{tabular}

\vspace{5mm}

The benchmarking reveals that when just taking into account the AST generation, the Rust implementation generally offers superior raw performance, but suffers from only currently supporting a subset of the full OpenQASM 3.0 specification. The QASM-TS parser provides competitive performance for web deployment scenarios, while the ANTLR parser offers a balance of features and performance suitable for development and testing.

\section{Acknowledgements}

We would like to thank Dr. Shohini Ghose for support and helpful discussions regarding the previous version of this software package, QASM-TS 1.0, which has had use in the community by hundreds (counted by npm downloads).

\bibliographystyle{acm}
\bibliography{main}

\end{document}